\documentclass[10pt,conference,compsoc]{IEEEtran}

\usepackage[nocompress]{cite} 

\usepackage{color,xcolor}
\usepackage[pdftex]{graphicx} 
\graphicspath{{./}}

\usepackage{amsmath}
\usepackage{amssymb}

\usepackage{threeparttable}
\usepackage{array}
\usepackage{multirow}

\usepackage[caption=false,font=footnotesize,labelfont=sf,textfont=sf,font=scriptsize]{subfig} 

\usepackage{enumitem}

\newcommand{\fcr}{$\square$\hspace*{-6.4pt}\raisebox{.8pt}{\color{gray}$\times$}}
\newcommand{\tcr}{$\square$}
\newcommand{\bcr}{\raisebox{.8pt}{\color{gray}$\times$}}

\newcommand{\Internet}{{\tt INTERNET}}
\newcommand{\CoarLocation}{{\tt ACCESS\_COARSE\_LOCATION}}
\newcommand{\FineLocation}{{\tt ACCESS\_FINE\_LOCATION}}
\newcommand{\Camera}{{\tt CAMERA}}
\newcommand{\Bluetooth}{{\tt BLUETOOTH}}
\newcommand{\ReadExtStorage}{{\tt READ\_EXTERNAL\_STORAGE}}
\newcommand{\WritExtStorage}{{\tt WRITE\_EXTERNAL\_STORAGE}}
\newcommand{\ReadContacts}{{\tt READ\_CONTACTS}}
\newcommand{\WritContacts}{{\tt WRITE\_CONTACTS}}
\newcommand{\MediaCtrl}{{\tt MEDIA\_CONTENT\_CONTROL}}
\newcommand{\ReadSMS}{{\tt READ\_SMS}}
\newcommand{\RecvSMS}{{\tt RECEIVE\_SMS}}
\newcommand{\SendSMS}{{\tt SEND\_SMS}}
\newcommand{\CallPhone}{{\tt CALL\_PHONE}}
\newcommand{\RecdAudio}{{\tt RECORD\_AUDIO}}
\newcommand{\Vibrate}{{\tt VIBRATE}}

\DeclareMathAlphabet{\mts}{U}{rsfs}{m}{n}
\DeclareMathOperator{\sgn}{sgn}

\newcommand{\bmt}{\boldsymbol}

\begin{document}
\IEEEoverridecommandlockouts

\thispagestyle{empty}

\begin{figure*}[!t]\large
This paper is a preprint; it has been accepted for publication in 16th IEEE International Conference on Dependable, Autonomic and Secure Computing --- DASC 2018 (DOI: 10.1109/DASC/PiCom/Data Com/CyberSciTec.2018.00127).
\medskip

{\bf IEEE copyright notice}
\smallskip

\copyright\ 2018 IEEE. Personal use of this material is permitted. Permission from IEEE must be obtained for all other uses, in any current or future media, including reprinting/republishing this material for advertising or promotional purposes, creating new collective works, for resale or redistribution to servers or lists, or reuse of any copyrighted component of this work in other works.
\vspace*{300pt}

\mbox{~}
\end{figure*}

\newpage

\title{WiP: Are cracked applications really free? An empirical analysis on Android devices\thanks{This work was supported by CYBER-TRUST project, which has received funding from the European Union's Horizon 2020 research and innovation programme under grant agreement no. 786698.}}

\author{\IEEEauthorblockN{%
Konstantinos-Panagiotis Grammatikakis\IEEEauthorrefmark{2},
Angela Ioannou\IEEEauthorrefmark{3},
Stavros Shiaeles\IEEEauthorrefmark{1} and
Nicholas Kolokotronis\IEEEauthorrefmark{2}%
\vspace*{8pt}}

\IEEEauthorblockA{\IEEEauthorrefmark{2}Department of Informatics and Telecommunications, University of Peloponnese, 22131 Tripolis, Greece\\Email: kpgram@uop.gr, nkolok@uop.gr}
\IEEEauthorblockA{\IEEEauthorrefmark{3}School of Pure and Applied Sciences, Open University of Cyprus, Nicosia 2220, Cyprus\\Email: angela.ioannou@st.ouc.ac.cy}
\IEEEauthorblockA{\IEEEauthorrefmark{1}Centre for Security, Communications and Networks Research, Plymouth University, Plymouth PL4 8AA, UK\\Email: stavros.shiaeles@plymouth.ac.uk}
}

\maketitle

\begin{abstract}
Android is among the popular platforms running on millions of smart devices, like smartphones and tablets, whose widespread adoption is seen as an opportunity for spreading malware. Adding malicious payloads to cracked applications, often popular ones, downloaded from untrusted third markets is a prevalent way for achieving the aforementioned goal. In this paper, we compare 25 applications from the official and third-party application stores delivering cracked applications. The behavioral analysis of applications is carried out on three real devices equipped with different Android versions by using five indicators: requested permissions, CPU usage, RAM usage and the number of opened ports for TCP and HTTP. Based on these indicators, we compute an {\em application intention score} and classify cracked applications as malicious or benign. The experimental results show that cracked applications utilize on average more resources and request access to more (dangerous) permissions than their official counterparts.
\end{abstract}

\section{Introduction}
\label{sec:intro}


This decade, we saw the development of more powerful and compact computing devices, like mobile phones, tablets, ultra portable computers and home appliances, whose capabilities were once possible only on personal computers. At the same time, the high degree of network connectivity and the provisioning of high-speed broadband services led to the development of novel services that take advantage of these new capabilities, hence transforming these common devices into {\em smart devices}. According to the Pew Research Center, $77\%$ of the total surveyed population in the United States owns a smartphone and $53\%$ owns a tablet \cite{pew18}. In addition to the traditional role of smart devices as a medium of communication and media consumption, with the advent of constant Internet connection capabilities many traditional services, such as financial or remote administration services, are now being carried out through mobile platforms. The popularity of the Android {\em operating system} (OS) grew with the popularity of smart devices, capturing the largest share of the mobile computing devices market, $74.39\%$ according to \cite{statcounter18b}, while also leading the personal computing devices market as a whole with a share of $40\%$ \cite{statcounter18a}. This widespread adoption of smart devices and the Android OS in particular, along with the increasing value of the services and applications provided, makes them a valuable target for malicious actors.

There are millions of applications available for the Android OS through the official application store (estimated around $3.5$ mil \cite{statista18}) and third-party application stores. The fact that the Android OS allows the installation of applications coming from third-party markets or other generally untrusted sources, also known as {\em sideloading}, in conjunction with the sheer number of available applications, makes malware installation a viable attack vector.

In this paper we use the term {\em malware} in order to denote malicious software, which is a type of computer programs, or applications in our case, developed with the intention to harm a computer network, system or its users \cite{sikorski12}. Some common malicious actions are related to abusing services, such as cellular data connections or SMS messages, displaying unwanted advertisements, facilitating the formation of botnets to launch large-scale attacks and disclosing sensitive or personal data to third parties. The legitimate applications whose digital rights management or copy protection controls have been removed are referred to as {\em cracked} applications; such versions of legitimate applications quite often carry malicious payloads. The Google Play store is the {\em official application store}, where the Android applications of most software vendors are distributed from. Further, applications obtained from the official application store are assumed to be benign and are next referred to as {\em official} applications when compared with their cracked counterparts.

The goal and motivation of this paper is to provide further insight into the price that users of smart devices actually pay by downloading cracked applications via unofficial, untrusted third-party stores. Towards this direction a sample of $25$ applications is used, where the official and cracked versions are compared against a number of indicators (permissions, CPU and RAM usage, as well as, open TCP and HTTP ports). Although an analysis of permissions requested by Android applications has been conducted in previous works, it is known that they cannot alone accurately identify malicious intent. Therefore, we study the extend to which the combined indicators could considerably increase the accuracy of classifying a cracked application as benign or malicious. Our analysis is carried out on the three most popular versions of the Android~OS: KitKat, Lollipop and Marshmallow. We observe that cracked applications request for more permissions, where the extra permissions are linked to malicious behavior, in addition to a tendency for utilization of more resources than the official applications. Moreover, our analysis illustrates that although newer versions of the Android OS are more efficient in resource management (CPU and RAM usage), the differences between cracked and official applications in these indicators are noticeable. In conjunction with the number of open TCP and HTTP ports, the set of indicators succeeds in efficiently delivering increased detection of malicious intent.

The rest of the paper is organized as follows: related work in the area of Android malware analysis is given in Section 2, whilst the design of the experimental process is presented in Section 3. The main findings of our analysis and concluding remarks are provided in Sections 4 and 5 respectively.

\section{Related work}
\label{sec:related}

The detection and analysis of malware on mobile devices has been an area of highly intensive research since their first appearance. The earliest attempts to create anti-malware systems were based on the installation of an {\em agent} on the mobile device, responsible for monitoring device activities and for reporting them to a central system to be further analyzed. One early example of such system was presented by \cite{cheng07}; an {\em agent} gathered communications data with to detect possible abuse of the communication capabilities of the mobile device. A similar system, proposed by \cite{houmansadr11} requires the {\em agent} to mirror the state of the device and its communications on a cloud-based central system to be emulated and further analyzed. A system using crowd sourcing to gather application activity data was presented by \cite{burguera11}. A signature-based agent system was presented by \cite{talha15}; the calculation of an {\em application malware score} (AMS) was performed on the central system by summing the {\em permission malware scores} (PMS) ---calculated from official store applications and known malware--- of each requested permission.

A behavioral analysis approach was taken by \cite{bose08} whose malware detection system relies on signatures generated by monitoring of the actions performed by the suspect application ---via the system events and {\em application programming interface} (API) calls--- and the construction of a logical flow diagram. Moreover, an analysis of $46$ malware samples and an evaluation of existing anti-malware solutions for the Android, iOS and Symbian platforms was performed by \cite{felt11a}. It is interesting to note that none of the iOS malware samples were approved by Apple's {\em App store}, indicating the need and effectiveness of human review.

In addition to the above lines of research, a permission-based analysis of mobile applications for detecting malware has been proposed. Towards this direction, $940$ applications were examined in \cite{felt11b} to determine whether the principle of {\em least privilege} had been followed. It was found that nearly one third of the Android applications violated the aforementioned principle, something that was attributed to developers misunderstanding the use of permissions and to the lack of a clear API documentation. An extension to the Android security enforcement system to also consider the relationship between the requested permissions was proposed by \cite{tang11}. This is justified by the fact that although individual permissions cannot indicate malicious intent, their relationship can be used to classify an application as malicious. A study of $125,229$ benign and malicious applications was carried out by \cite{huang13} using the requested permissions as an indicator of intent. The performance of four machine learning algorithms in terms of their detection rate draw the conclusion that although permissions alone can be used to quickly classify applications as malicious or benign, a secondary analysis is required to make the final decision. A more extensive analysis of an application's manifest file ---based on searching for terms pertaining to permission requests, process names and identifiers--- was proposed by \cite{sato13}. Likewise, they concluded that textual analysis of the manifest file is resource efficient and when combined with other techniques it can improve the accuracy of the analysis. Many works rely on binary classification to decide whether an application is malicious or benign; although in certain cases a higher granularity would be needed for providing a more accurate characterization of an application. To solve this problem, \cite{gates14} proposed the use of {\em risk scoring functions} to calculate an overall value that is subsequently used to characterize an application. A permission-based system was presented by \cite{li18} considering only permissions that are rarely requested by malicious or benign applications and using machine learning to differentiate between the two classes.

Along with the development of anti-malware systems, the problem of classifying malware and security threats for mobile platforms has also been considered. An analysis of $1,260$ Android malware samples was conducted by \cite{zhou12} using an evolutionary-based approach. It was found that about $86\%$ of the samples were repackaged and therefore they highlighted the need for reviewing the applications in Android application stores (not only the official one), just as \cite{felt11a} proposed. Quite recently, \cite{sadeghi17} conducted a more thorough review of the existing literature, from 2008 to 2016, where taxonomies on many different areas and approaches used in the literature for Android malware analysis were presented.

\section{Proposed methodology}
\label{sec:method}

In order to study the differences in the behavior between benign applications downloaded from the official application store and the cracked ones, a sample of $n = 25$ applications was used as a proof of concept ---that is, to demonstrate that cracked applications often carry malicious payloads with the intention to harm the mobile device where they are installed or its user. The applications in our sample, which are listed alphabetically in Table \ref{tab:apps}, were randomly selected from two third-party stores\footnote{Cracked applications were downloaded from {\tt CrackAPK.com} and {\tt AppCake.net}, which both accept user--uploaded applications.} and were tested on the three Android OS versions with the largest market share: KitKat, Lollipop and Marshmallow. Instead of analyzing the behavior of the sample applications in a simulated environment, the setup of the experimental process involved using three Samsung Galaxy mobile devices
\begin{itemize}
\item S3 neo with Android v4.4.2 (KitKat),
\item S5 with Android v5.0 (Lollipop), and
\item S7 with Android v6.0 (Marshmallow)
\end{itemize}
in order to study the applications' behavior in real life use case scenarios and avoid detection of the simulated process by potential malicious payloads; they often exhibit different behavioral patterns if simulated environments are detected.

\begin{table}[h]
\begin{threeparttable}[t]
\setlength{\tabcolsep}{5.8pt}
\renewcommand{\arraystretch}{1.0}
\centering
\caption{List of application samples.}
\label{tab:apps}
\begin{tabular}{ @{}r | l@{} c @{}r | l@{} }
	\cline{1-2}\cline{4-5}
	         \#\, & Name                        &\hspace*{6pt}& \#\, & Name \\
	\cline{1-2}\cline{4-5}
	 1\tnote{a}\, & 360 Security -- Antivirus Free & & 14\tnote{a}\, & Lunchbox \\
	 2\tnote{b}\, & 3C Toolbox                     & & 15\tnote{a}\, & Mean Spheres Attack \\
	 3\tnote{a}\, & 3D Charts                      & & 16\tnote{b}\, & Mobile Security \& Antivirus \\
	 4\tnote{a}\, & 4Shared                        & & 17\tnote{a}\, & Piques \\
	 5\tnote{b}\, & 9GAG                           & & 18\tnote{a}\, & Root Browser \\
	 6\tnote{a}\, & Audio Manager                  & & 19\tnote{b}\, & Root Tool Case \\
	 7\tnote{a}\, & Calorie Counter                & & 20\tnote{a}\, & Run Cow Run \\
	 8\tnote{a}\, & Clean Master -- Free Antivirus & & 21\tnote{b}\, & Smart IPTV \\
	 9\tnote{b}\, & Dual Sim Selector              & & 22\tnote{a}\, & Solar System Scope \\
	10\tnote{a}\, & Enemy Strike                   & & 23\tnote{a}\, & Spy Mouse \\
	11\tnote{a}\, & Fillshape                      & & 24\tnote{b}\, & Unit Converter \\
	12\tnote{a}\, & John NES -- NES Emulator       & & 25\tnote{a}\, & Vector \\
	13\tnote{a}\, & Link2SD                        & &               & \\
	\cline{1-2}\cline{4-5}
\end{tabular}
\begin{tablenotes}
\item [a] Obtained from the {\em official} store and {\tt CrackAPK.com}
\item [b] Obtained from the {\em official} store and {\tt AppCake.net}
\end{tablenotes}
\end{threeparttable}
\end{table}

Our approach is based on assumptions about the behavioral patterns exhibited by applications. In particular, official applications are considered to be benign by default, as their intentions are stated on their official store and consent is given at install-time; the same holds with cracked applications, which are assumed to be benign unless proven otherwise. Cracked applications displaying significant behavioral deviations from the official ones are considered to be malicious. It is noted that the differences in an indicator alone may not suggest malicious behavior, since small differences could be well attributed to the deletion or insertion of bytecode by the patching process; on the other hand, deviations in many indicators increase the possibility of malicious intentions. The requested permissions exhibit varying degrees of correlation to malicious intent ---as evidenced in the literature \cite{felt11a,gates14,li18,zhou12}--- and their study can classify an application as malicious or benign.

Malware analysis is usually performed either manually by a security analyst, or automatically by special software, and there are three prevalent approaches: static, dynamic and hybrid \cite{sikorski12,sadeghi17}. Static methods focus on characteristics such as an application's binary code, structure and metadata, while dynamic approaches aim at analyzing an application's behavior during its execution. Hybrid techniques combine both static and dynamic aspects to get a more complete view of the suspect application; our approach can be classified as such, since we measure both static and dynamic indicators. Due the need for analyzing both the official and the cracked version of each application on three Android OSs, we need to choose indicators that may be measured accurately and efficiently. In particular, the indicators used in our study for each application $i = 1,\ldots,n$ are the requested permissions $\bmt{p}_i$, CPU usage $c_i$, RAM usage $r_i$, as well as the number of ports opened for TCP and HTTP communications that are denoted by $t_i$ and $h_i$ respectively.

The first indicator, the requested permissions $\bmt{p}_i$, can be obtained by the application manifest file. A total of $m = 16$ permissions were tracked, which are listed in Table \ref{tab:permissions.list}, being the union of those requested by the applications; thus we let $\bmt{p}_i = (p_{i1},\ldots,p_{im})$, where $p_{ij} \in \{0,1\}$ indicates if the $j$th permission is requested by the $i$th application. Permissions are requested by an application to obtain access to hardware resources, e.g.\ the microphone or camera, and to restricted API functions by declarations in the manifest file \cite{elenkov15}. They are granted at install-time or at run-time (from version 6.0, Marshmallow and later) and are classified in three protection levels \cite{android18}: {\em normal}, {\em dangerous} and {\em special}. The manifest file {\tt AndroidManifest.xml} is found inside the {\em Android package} (APK) file, which constitutes the main way that all applications are distributed and installed on the Android OS. We used the {\tt Show Java} application on the mobile devices to unpack and extract the files contained in APK files.

\begin{table}[h]
\begin{threeparttable}[t]
\setlength{\tabcolsep}{3.8pt}
\renewcommand{\arraystretch}{1.0}
\centering
\caption{Measured permissions ({\tt android.permission.*}).}
\label{tab:permissions.list}
\begin{tabular}{ @{}r | l@{} c @{}r | l@{} }
	\cline{1-2}\cline{4-5}
	         \#\, & Name                       & \hspace*{6pt}& \#\, & Name \\
	\cline{1-2}\cline{4-5}
	 1\tnote{a}\, & \Internet                 & &  9\tnote{c}\, & \MediaCtrl \\
	 2\tnote{b}\, & \CoarLocation & & 10\tnote{b}\, & \ReadSMS \\
	 3\tnote{b}\, & \FineLocation   & & 11\tnote{b}\, & \RecvSMS \\
	 4\tnote{b}\, & \Camera                   & & 12\tnote{b}\, & \SendSMS \\
	 5\tnote{a}\, & \Bluetooth                & & 13\tnote{b}\, & \WritContacts \\
	 6\tnote{b}\, & \ReadExtStorage  & & 14\tnote{b}\, & \CallPhone \\
	 7\tnote{b}\, & \ReadContacts           & & 15\tnote{b}\, & \RecdAudio \\
	 8\tnote{b}\, & \WritExtStorage & & 16\tnote{a}\, & \Vibrate \\
	\cline{1-2}\cline{4-5}
\end{tabular}
\begin{tablenotes}
\item [a] Classified as {\em Normal}
\item [b] Classified as {\em Dangerous}
\item [c] Classified as {\em Special}
\end{tablenotes}
\end{threeparttable}
\end{table}

CPU and RAM usage measurements can be obtained by the Android OS application monitoring services (the means of access differs between Android OS versions). The usage of CPU and RAM may indicate differences in the bytecode or the memory consumption patterns between the official and a cracked version. The values reported by the Android OS on each mobile device were used.

The number of open TCP and HTTP ports was obtained by packet inspection of the network traffic generated by each application. The {\em hypertext transfer protocol} (HTTP) is often utilized by malware for communicating with a {\em command and control} server to receive new commands for data extraction or to download files on the infected devices. It is commonly used by legitimate applications to download resources and to use APIs available through the Internet. This makes less suspicious the use of HTTP and in addition HTTP traffic is widely allowed to pass through the network firewalls. The {\em transport control protocol} (TCP) ---which is widely used for providing, at layer 4 of the {\em open systems interconnection} (OSI) model, connection-oriented and reliable data stream services that an application requires for sending and receiving error--free data--- was monitored to capture suspicious connections established by malicious payloads having been included in cracked applications. We have used ``Wireshark'' on a computer on the same network with the mobile devices to capture and analyze the generated network traffic.

\subsection{Application intention score}
\label{sec:score}

Based on the above, an application is characterized by~a tuple $a_i = (\bmt{p}_i, c_i, r_i, t_i, h_i)$, $i=1,\ldots,n$, where $a_i^o$ and~$a_i^c$ are used to differentiate between the official and the cracked versions' profiles. Since permissions exhibit varying degrees of correlation to malicious intent \cite{felt11a,gates14,li18,zhou12}, a number of $k = 3$ permission groups $\Pi_l$, for $l=1,\ldots,k$, were defined to simplify the analysis. Group $\Pi_1$ contains the permissions considered to be highly indicative of malicious behavior, i.e. those in the set $\{1,10,\ldots,16\}$. The group $\Pi_2$ includes the permissions $\{6,7,8\}$ that could suggest malicious intention and have a smaller correlation compared to the permissions in the first group. Finally, the group $\Pi_3$ has the remaining permissions $\{2,\ldots,5,9\}$ that are commonly requested from both malicious and benign applications. As in \cite{gates14}, we define a mapping that provides an overall value characterizing an application's intentions; this is called {\em application intention score} $s \in [-1,1]$ and is determined by
\begin{equation}\label{equ:intention}
	s_{i} = \sum_{l=1}^k w_l \, \delta_{il}, \quad i = 1, \ldots, n
\end{equation}
where $w_l$ is the weight assigned to the permission group $\Pi_l$, with $w_1 + \cdots + w_k = 1$, and $\delta_{il} \in \{-1, 0, 1\}$. The term $\delta_{il}$ represents the {\em group difference score} of the $i$th application%
\begin{equation}\label{equ:difference}
	\delta_{il} = \sgn\left( \sum_{j \in \Pi_l} \bigl( p_{ij}^o - p_{ij}^c \bigr) \right)
\end{equation}
where $\sgn(\cdot)$ is the signum function for which we have that $\sgn(0) = 0$ by convention. Note that $\delta_{il} < 0$ if and only if the cracked application requests more permissions than the official one. Moreover, we define $\mts{L}:[-1,1] \rightarrow L$ as
\begin{equation}\label{equ:classes}
\mts{L}(s) = \begin{cases}
\ell_1: \text{``malicious''}\,, & \text{if $s < -0.4$} \\
\ell_2: \text{``rather malicious''}\,, & \text{if $-0.4 \leq s < 0$} \\
\ell_3: \text{``rather benign''}\,, & \text{if $0 \leq s \leq 0.4$} \\
\ell_4: \text{``benign''}\,, & \text{otherwise}
\end{cases}
\end{equation}
mapping the application intention score onto a set of classes or labels $L = \{\ell_1,\ldots,\ell_4\}$ characterizing cracked apps with respect to the difference in the requested permissions. Using this classification, we next seek for correlation with the other indicators measured in this analysis.

\section{Experimental results}
\label{sec:results}

In this section, we present the results of our analysis for the sample applications used. The requested permissions per application (official and cracked ones) are listed in Table \ref{tab:permissions.detailed}.

\begin{table}[h]
\begin{threeparttable}[t]
\setlength{\tabcolsep}{3.1pt}
\renewcommand{\arraystretch}{1.0}
\centering
\caption{Permissions requested per application.}
\label{tab:permissions.detailed}
\begin{tabular}{r|cccccccccccccccc}
\hline
      & \multicolumn{16}{c}{{\tt anrdoid.permission.*}} \\
\,App & 1 & 2 & 3 & 4 & 5 & 6 & 7 & 8 & 9 & 10 & 11 & 12 & 13 & 14 & 15 & 16 \\
\hline
 1 & \fcr & \fcr & \fcr & \fcr & \fcr & \fcr & \fcr & \bcr & & \bcr & \bcr & \bcr & \bcr & \bcr & & \\
 2 & \fcr & \fcr & \fcr & \fcr & & & \fcr & \tcr & \fcr & \bcr & \bcr & \bcr & & & & \\
 3 & \bcr & \bcr & \bcr & \bcr & & \bcr & & \bcr & \tcr & & & & & & \bcr & \bcr\\
 4 & \fcr & \fcr & \fcr & \bcr & & \bcr & \tcr & \bcr & \tcr & & & & & & \bcr & \bcr\\
 5 & \fcr & & & & & \bcr & & \bcr & \fcr & & & & & & & \\
 6 & \fcr & \fcr & \fcr & \fcr & & \bcr & & \bcr & \tcr & & & & & & \bcr & \bcr\\
 7 & \bcr & \bcr & \bcr & \fcr & & \bcr & & \bcr & \tcr & & & & & & \bcr & \bcr\\
 8 & \fcr & & & \fcr & & \bcr & & \bcr & \fcr & & & & & & & \\
 9 & & & & & & \bcr & \fcr & \bcr & \fcr & & & & & & & \\
10 & \fcr & \bcr & \bcr & \bcr & & \bcr & \tcr & \bcr & \tcr & & & & & & \bcr & \bcr\\
11 & \bcr & \bcr & \bcr & \bcr & & \bcr & & \bcr & \tcr & & & & & & \bcr & \bcr\\
12 & \fcr & \bcr & \bcr & \bcr & & \bcr & & \bcr & \tcr & & & & & & \bcr & \bcr\\
13 & \bcr & \bcr & \bcr & \bcr & & \bcr & & \bcr & \tcr & & & & & & \bcr & \bcr\\
14 & \fcr & \bcr & \bcr & \bcr & & \bcr & & \bcr & \tcr & & & & & & \bcr & \bcr\\
15 & \bcr & \bcr & \bcr & \bcr & & \bcr & & \bcr & \tcr & & & & & & \bcr & \bcr\\
16 & \fcr & \fcr & \fcr & \fcr & & \bcr & \fcr & \bcr & \fcr & \bcr & \bcr & \bcr & & & & \\
17 & \bcr & \bcr & \bcr & \bcr & & \bcr & & \bcr & \tcr & & & & & & \bcr & \bcr\\
18 & \fcr & \bcr & \bcr & \bcr & & \bcr & & \bcr & \tcr & & & & & & \bcr & \bcr\\
19 & \bcr & \bcr & \bcr & & & & & & \fcr & & & & & & & \\
20 & \fcr & \bcr & \bcr & \bcr & & \fcr & & \fcr & & & & & & & \bcr & \bcr\\
21 & \fcr & \bcr & \bcr & & & & & & & & & & & & & \\
22 & \bcr & \fcr & \fcr & \bcr & & \bcr & & \bcr & & & & & & & \bcr & \bcr\\
23 & \fcr & \fcr & \fcr & \bcr & & \bcr & & \bcr & & & & & & & \bcr & \bcr\\
24 & \fcr & \bcr & \bcr & & & & & & & & & & & & & \\
25 & \fcr & \bcr & \bcr & \bcr & & \bcr & & \bcr & \tcr & & & & & & \bcr & \bcr\\
\hline
\multicolumn{17}{l}{\rule{0pt}{12pt}Overall} \\
\hline
\tcr\, & 16 & 7 & 7 & 6 & 1 & 2 & 6 & 2 & 19 & 0 & 0 & 0 & 0 & 0 & 0 & 0 \\
\bcr\, & 24 & 22 & 22 & 20 & 1 & 21 & 4 & 21 & 6 & 3 & 3 & 3 & 1 & 1 & 16 & 16 \\
\hline
\end{tabular}
\begin{tablenotes}
\item [] \hspace*{-6pt}\tcr\ Requested by the {\em official} application
\item [] \hspace*{-6pt}\bcr\ Requested by the {\em cracked} application
\item [] \hspace*{-6pt}\fcr\ Requested by both application types
\end{tablenotes}
\end{threeparttable}
\end{table}

The weights that have been empirically assigned in \eqref{equ:intention} are equal to $w_{1} = 0.6$, $w_{2} = 0.3$, and $w_{3} = 0.1$. By computing the application intention score, the cracked applications of Table \ref{tab:apps} are classified according to the mapping $\mts{L}$, and the results obtained are as follows:
\begin{itemize}
	\item $\ell_1$ contains 1, 3--4, 6--7, 10--18, 22--23, and 25.
	\item $\ell_2$ contains 2, 19, and 20.
	\item $\ell_3$ contains 5, 8, and 9.
	\item $\ell_4$ contains 21 and 24.
\end{itemize}
In general, we see that the cracked applications tend to request more permissions than the official applications, with an overall average $7.36$ versus $2.64$ permissions. We note that nineteen cracked applications requested permissions 10--16 and eight cracked applications requested Internet access (permission 1) even though their official versions did not; all of them are classified as malicious or rather malicious. Also permissions related to SMS messaging (10--12) were requested together as a set; this pattern was also observed with permissions 6 and 8 that are related to read/write access to external storage.

\begin{figure}[t]
\centering
\includegraphics[width=\linewidth]{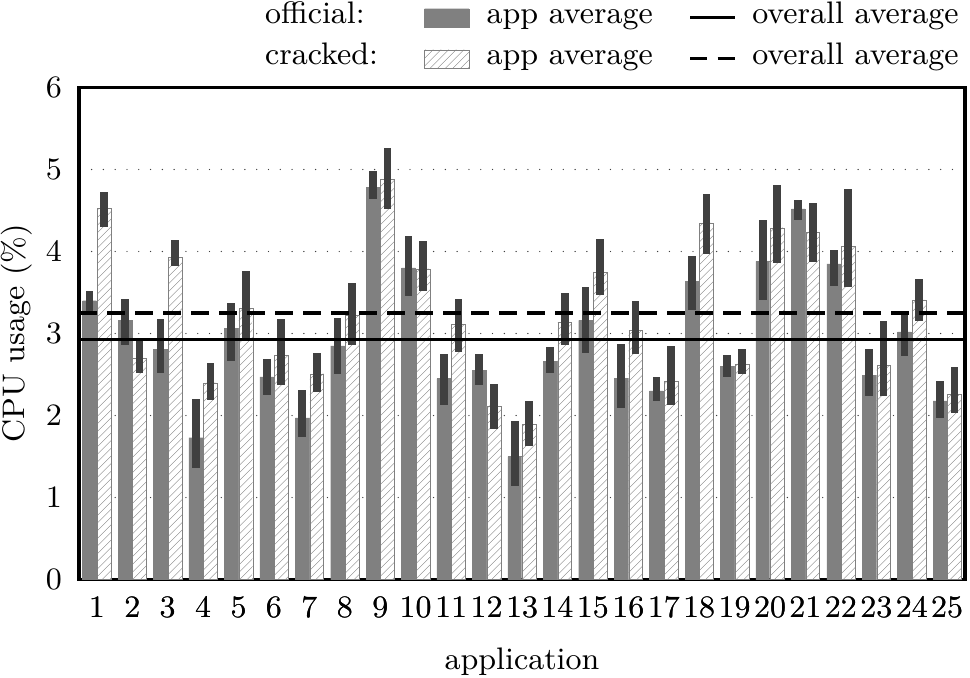}
\caption{The usage of CPU per application.}
\label{fig.cpu.app.detailed}
\end{figure}

\begin{figure}[t]
\centering
\includegraphics[width=\linewidth]{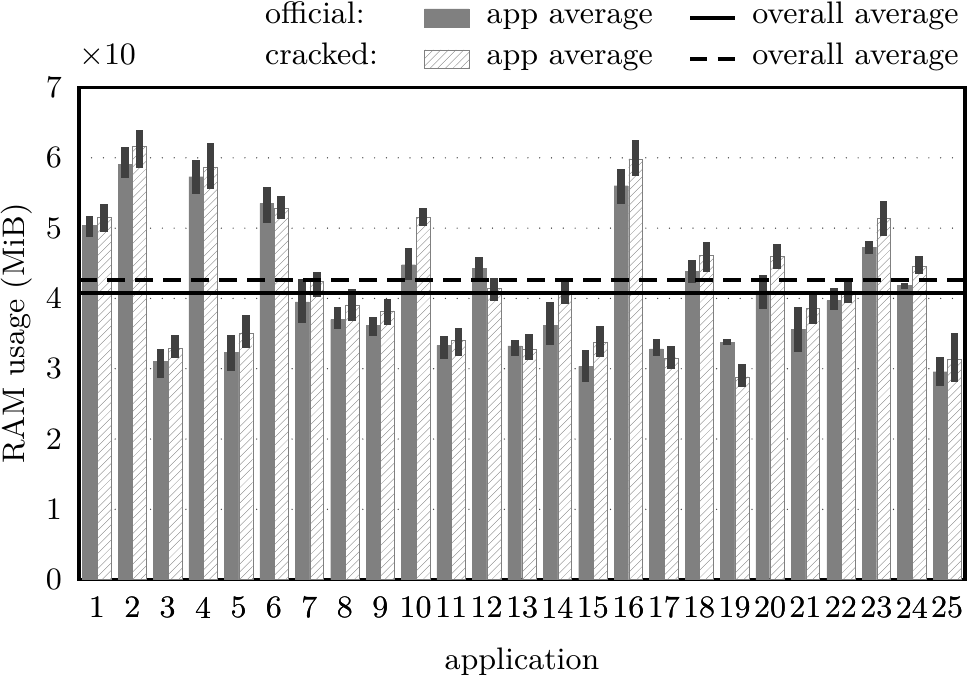}
\caption{The usage of RAM per application.}
\label{fig.ram.app.detailed}
\end{figure}

The measured CPU usage (\%) and the measured RAM usage (MiB) across all Android OS versions are presented in Figures \ref{fig.cpu.app.detailed} and \ref{fig.ram.app.detailed} respectively. In general, cracked applications tend to utilize slightly more CPU and RAM resources than their official counterparts. The overall CPU usage average is $3.25\%$ for cracked applications in contrast to $2.93\%$ for the official applications; moreover, the overall RAM usage average is $42.65$ MiB and $40.80$ MiB for the cracked and the official applications respectively. In all figures, the box plots (\raisebox{.5pt}{\color{darkgray}\rule{7pt}{3.5pt}}) indicate the minimum and maximum values among the Android versions considered. We noticed that, with very few exceptions, the maximum (resp. minimum) values for both CPU and RAM usage were attained on KitKat (resp. Marshmallow) implying that possible use of these indicators in newer versions of the Android OS is rather hard due to the more efficient use of the available resources.

\begin{figure}[t]
\centering
\includegraphics[width=\linewidth]{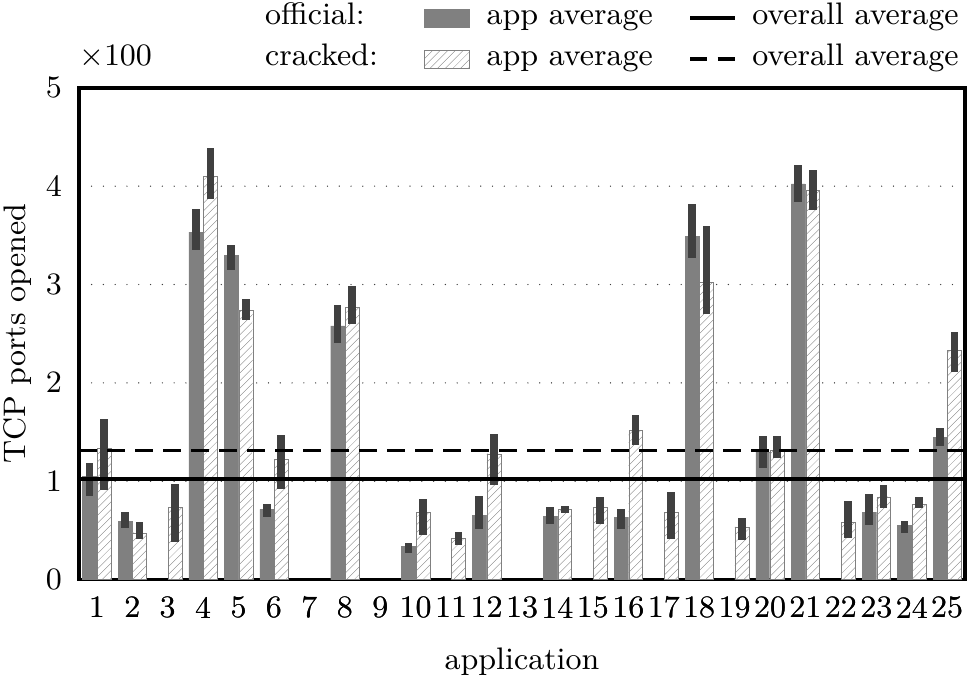}
\caption{The number of TCP ports opened per application.}
\label{fig.tcp.app.detailed}
\end{figure}

\begin{figure}[t]
\centering
\includegraphics[width=\linewidth]{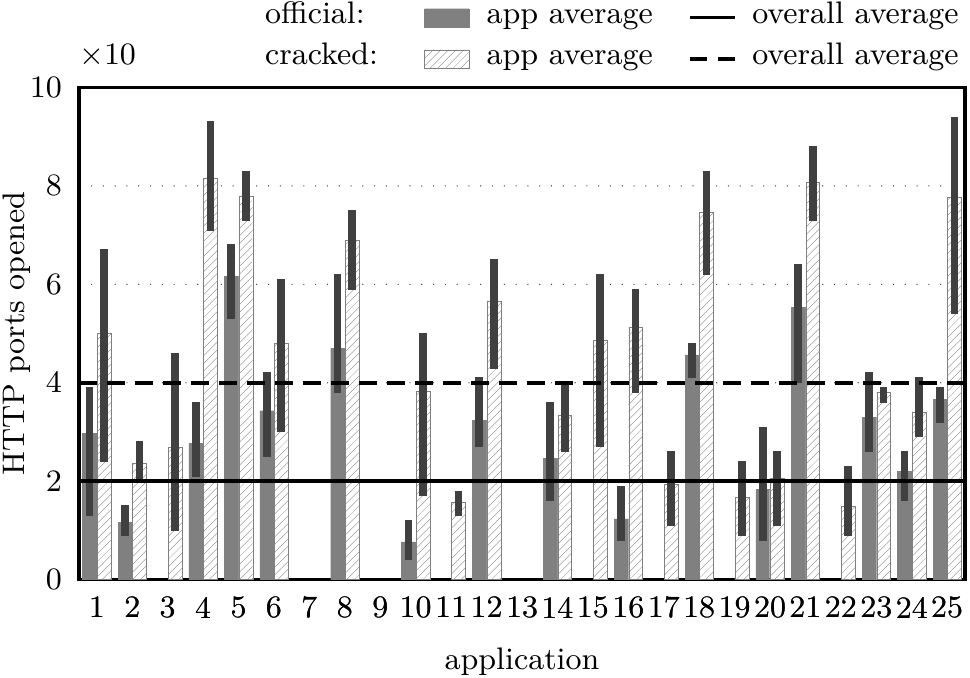}
\caption{The number of HTTP ports opened per application.}
\label{fig.http.app.detailed}
\end{figure}

The number of the open ports for TCP and HTTP per application across all Android OS versions are presented in Figures \ref{fig.tcp.app.detailed} and \ref{fig.http.app.detailed} respectively. Clearly, cracked applications in most cases open more ports for both protocols than official applications. The overall average of the TCP ports opened is $131.19$ and $102.15$ for cracked and official applications respectively, whereas the overall average of the HTTP ports opened is $39.92$ and $20$ for cracked and official applications respectively.

The average usage overheads that are measured by using cracked applications, across all the Android OS versions, are presented in Table \ref{tab:classification.avg} for each application class. We observe that cracked applications, classified as malicious according to \eqref{equ:classes}, in most cases utilize significantly more resources and request more permissions linked to malicious behavior ({\em see} group $\Pi_1$) than cracked applications having been classified as benign; this illustrates the existence of a clear difference between these two extreme ends. Furthermore, the overhead incurred by cracked applications classified as rather benign/ malicious confirm the uncertainty of our classification; more precisely, rather malicious apps utilize less CPU and RAM resources, whereas rather benign apps generate significantly less network traffic as evidenced by the additional TCP and HTTP ports opened. These differences seem to confirm the existence of correlation between the classification used and and the four new indicators. However, due to the fact that the applications' sample was small ---just used to establish a proof of concept--- further and more extensive testing of cracked applications should be performed to prove  positive impact in the detection of malware for the Android OS.

\begin{table}[t]
\setlength{\tabcolsep}{5pt}
\renewcommand{\arraystretch}{1.0}
\centering
\caption{Average usage overhead per application class.}
\label{tab:classification.avg}
\begin{tabular}{r|rrrr}
	\hline
	& malicious & rather malicious & rather benign & benign\\
	\hline
	 CPU  (\%) &  $0.43$\hspace*{6pt} & $-0.01$\hspace*{16pt} &   $0.24$\hspace*{9pt} &  $0.06$\hspace*{1pt} \\
	 RAM (MiB) &  $1.81$\hspace*{6pt} &  $0.91$\hspace*{16pt} &   $2.03$\hspace*{9pt} &  $2.84$\hspace*{1pt} \\
	TCP  ports & $41.29$\hspace*{6pt} & $14.78$\hspace*{16pt} & $-12.11$\hspace*{9pt} &  $8.00$\hspace*{1pt} \\
	HTTP ports & $23.02$\hspace*{6pt} & $10.34$\hspace*{16pt} &  $12.78$\hspace*{9pt} & $18.67$\hspace*{1pt} \\
	\hline
\end{tabular}
\rule{0pt}{10pt}{\em Positive values indicate increased resource usage by cracked applications}
\end{table}

\section{Concluding remarks}
\label{sec:conclusion}

An empirical analysis of cracked applications running on Android platforms was carried out in this paper. The sample set consisted of $25$ applications whose cracked and official versions were compared against a number of indicators: the requested permissions, CPU usage, RAM usage, as well as, the number of ports opened for TCP and HTTP. Following the introduction of an application intention score function, which relies on the permissions requested by the application, cracked applications were classified into groups associated with varying likelihoods of carrying malicious payloads. The extent at which the information provided by other indicators can increase the accuracy of classification is considered. 

Although any deviations in CPU and RAM usage (resp.\ in the number of TCP and HTTP ports opened) alone are often not indicative of malicious behavior, when paired with reliable malware detection methods their accuracy can be considerably increased. Our preliminary results across all the tested Android OS versions show that cracked applications request on average more permissions, tend to utilize slightly more CPU and RAM resources and open more TCP and HTTP ports than official applications; the classification resulted in about $80\%$ of the cracked applications to be classified as malicious or rather malicious. These findings suggest that cracked applications have questionable intentions, that users should be vigilant when installing cracked and untrusted applications, and that human review is required in both official and third-party application stores.

Possible directions for future work include the increase of the sample size in order to obtain statistically robust results and yield more accurate information about how permissions and values in the rest of the indicators are distributed for each application type (cracked and official ones). Differences in the distributions can further be used to design accurate application intention score functions and help detect malicious payloads in cracked applications.


\end{document}